\documentclass[12pt]{article}

\usepackage{newtxtext,newtxmath}

\usepackage{graphicx}
\usepackage{ tipa }
\usepackage{float}

\usepackage[letterpaper,margin=1in]{geometry}

\renewenvironment{abstract}
	{\quotation}
	{\endquotation}

\date{}

\makeatletter
\renewcommand{\fnum@figure}{{Figure \thefigure}}
\renewcommand{\fnum@table}{{Table \thetable}}
\makeatother

\usepackage{scicite}

\usepackage{url}

\usepackage{lineno}

\def\scititle{
From Gas to Ice Giants: A Unified Mechanism for Equatorial Jets}
\title{\bfseries \boldmath \scititle}

\author{
	Keren Duer-Milner$^{1,2,3\ast}$,
	Nimrod Gavriel$^{1}$,
	Eli Galanti$^{1}$,
        Eli Tziperman$^{4,5}$,
        Yohai Kaspi$^{1\ast}$\and
	\small$^{1}$Department of Earth and Planetary Sciences, Weizmann
Institute of Science, Rehovot, Israel.\and
	\small$^{2}$Leiden Observatory, Leiden University, Niels Bohrweg
2, 2333 CA Leiden, the Netherlands.\and
    \small$^{3}$SRON Netherlands Institute for Space Research, Niels
Bohrweg 4, 2333 CA, Leiden, the Netherlands.\and
    \small$^{4}$Department of Earth and Planetary Sciences, Harvard
University, Cambridge, Massachusetts.\and
    \small$^{5}$School of Engineering and Applied Sciences, Harvard
University, Cambridge, Massachusetts.\and
	\small$^\ast$Corresponding author. Email: duer@strw.leidenuniv.nl (K.D.~M.); yohai.kaspi@weizmann.ac.il (Y.K.)\and
    \\
    DOI: 10.1126/sciadv.ads8899
}

\begin{document} 

\maketitle

\begin{abstract} \bfseries \boldmath
The equatorial jets dominating the dynamics of the Jovian planets 
exhibit two distinct types of zonal flows: strongly eastward in the gas giants (superrotation)
and strongly westward in the ice giants (subrotation). Existing theories propose different
mechanisms for these patterns, but no
single mechanism has successfully explained both. However, the planetary
parameters of the four Solar System giant planets suggest that a fundamentally
different mechanism is unlikely. In this study, we show that convection-driven columnar structures can account for both eastward and westward equatorial jets, framing the phenomenon as a bifurcation. Consequently, both superrotation and subrotation emerge
as stable branches of the same mechanistic solution. Our analysis
of these solutions uncovers similarities in the properties of equatorial
waves and the leading-order momentum balance. This study suggests that the fundamental dynamics governing equatorial jet formation may be more broadly applicable across the Jovian planets than previously believed, offering a unified explanation for their two distinct zonal wind patterns.
\end{abstract}

\noindent

\subsection*{Teaser}

A bifurcation in equatorial jet formation can explain the differences between the gas and ice giants.

\section*{Introduction}

\begin{figure}[H]
\begin{centering}
\includegraphics[width=1\textwidth]{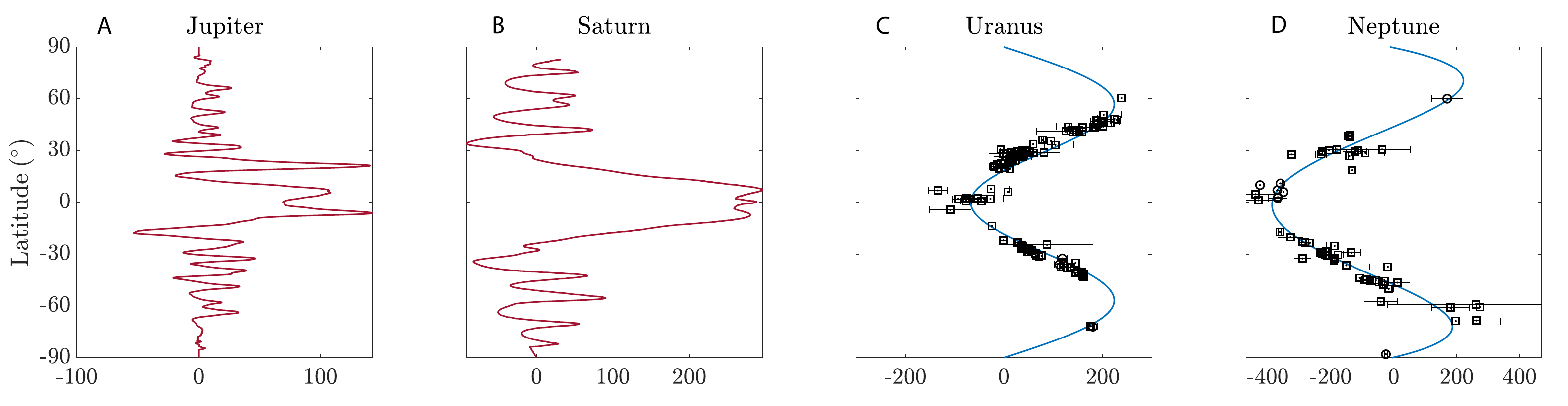}
\par\end{centering}
\caption{\label{fig:Observations}\textbf{Measurements of zonal wind profiles of the
Jovian planets.} Zonally averaged zonal cloud-level winds $\left[{\rm m\,s^{-1}}\right]$
of (A) Jupiter \cite{Tollefson2017}, (B) Saturn \cite{garcia2011},
(C) Uranus \cite{Hammel2001,Hammel2005,Sromovsky2005}, and (D) Neptune
\cite{Sromovsky1993,Sromovsky2001a}. Saturn's wind field incorporates
recent estimations of the planet's rotation rate \cite{mankovich2019}.
The winds on Uranus and Neptune were measured by Voyager 2 (circles)
and HST (squares), and are presented along with a polynomial fit of
the data (line).}
\end{figure}

The Jovian planets - Jupiter, Saturn, Uranus, and Neptune, exhibit
fascinating atmospheric dynamics, particularly evident in their zonal
wind patterns. One of the most distinguishing characteristics among them is the direction
of their equatorial jets \cite{Ingersoll1990}. Jupiter and Saturn
exhibit equatorial superrotation, meaning they flow prograde (west to east),
in the same direction as the planet\textquoteright s rotation. {The
term ``superrotation'' refers to wind velocity that exceeds the angular momentum conserving wind, which is defined using a simple conservation of angular momentum (eq.~\ref{eq:superrotation})
\cite{imamura2020}}. Jupiter's jet-stream reaches velocities
of $\sim 100$ {}m~s$^{-1}${} \cite{Tollefson2017} at the equator, and Saturn, while
similar in size and rotation to Jupiter, exhibits a broader and more
pronounced equatorial jet, with peak speeds of approximately 300 {}m~s$^{-1}${}
\cite{garcia2011,mankovich2019} (Fig.~\ref{fig:Observations}). 

In contrast, Uranus and Neptune have retrograde equatorial jets (east
to west), moving opposite to the direction of their rotation (subrotation).
The equatorial jet on Uranus reaches wind speeds of $\sim50\,{\rm m\,s^{-1}}$
\cite{Hammel2001,Sromovsky2005,Hammel2005} and Neptune's westward
jet reaches $\sim400\,{\rm m\,s^{-1}}$ at the equator \cite{Conrath1989,Sromovsky1993,Sromovsky2001a}. {Another notable feature is the presence of alternating midlatitude jets, which are prominent on Jupiter and Saturn, while Uranus and Neptune each have only one jet per hemisphere (Fig.~\ref{fig:Observations})}.
The wind velocity measurements are relative to a uniform rotation rate, which is associated with the planets' interior and aligned with their magnetic field \cite{desch1981}. This rotation rate is determined using radio emissions detected by various spacecraft, along with other methodologies  [e.g.,~\cite{helled2009b,Helled2010a,Helled2015,mankovich2019}].
Over the past few decades, the jets on all Jovian planets have shown
consistency, at least since the advent of modern measurement techniques.

{The difference in equatorial jet direction presents a major challenge in planetary science. While it has been suggested in previous studies that the planetary dynamics could be driven by different mechanisms, it is also possible that they are influenced by similar processes due to their comparable physical characteristics (tab.~S1)}. {All four planets have comparable rotation periods and penetration depths of zonal winds. Jupiter, Saturn, and Neptune also exhibit comparable normalized internal infrared flux. Although Uranus' flux is slightly smaller, its zonal wind structure is remarkably similar to Neptune's, as are their size and mass.} The jets on Jupiter penetrate approximately $0.95$ of its radius, according
to measured gravity field estimations \cite{Kaspi2018,duer2020} and ohmic
dissipation constraints \cite{Liu2008}. The jets on Uranus, and
Neptune, penetrate to a maximal depth of $\sim0.95$ \cite{Kaspi2013c,soyuer2020},
meaning they are shallower or equal to the jets on Jupiter (to date, the winds penetration depth of the ice giants has not yet been determined). Saturn's
jets reach about 0.84 of its radius \cite{galanti2019, kaspi2020},
indicating a larger proportion related to zonal jets. Other parameters,
such as obliquity - where Uranus’s is distinct at 98° - vary notably among the planets.
{However, the considerable differences between Uranus and Neptune are unlikely to be critical in determining the zonal wind mechanisms, as indicated by the similar zonal wind patterns observed on both ice giants. All of this suggests that a common mechanism may be responsible for jet formation and truncation at depth on the Jovian planets.}

Theories regarding the formation of zonal jets on giant planets focus
on two main mechanisms: convection driven by internal heating [e.g.,~\cite{Busse1994,Christensen2001,Aurnou2001,Heimpel2005,Kaspi2007,Jones2009,Kaspi2009,gastine2013}]
and baroclinic instability resulting from latitudinal variations in
solar radiation [e.g.,~\cite{Williams1978,Cho1996b,Cho1996}].
Studies utilizing a thin-shell general circulation model have demonstrated
that both superrotation and subrotation can occur depending on the
energy scheme \cite{Schneider2009,Liu2010,young2019,Lian2008,guendelman2025}. {Specifically,
solar forcing tends to produce retrograde equatorial jets as baroclinic
eddies transport momentum poleward, whereas strong intrinsic heat fluxes leads to prograde
equatorial jets due to equatorward momentum convergence by equatorial
waves \cite{Liu2010}. However, these variations may not adequately account for the differences between the planets, due to their physical characteristics (tab.~S1). Additionally, jets in thin-shell simulations are generally baroclinic and tend to vanish at depths of a few to $\sim 100$ bars, although they may penetrate deeper in a different numerical setup \cite{Schneider2009,Liu2010,guendelman2025} This somewhat contradicts recent estimates indicating that Jupiter's and Saturn's jets are relatively barotropic and extend deep ($\sim10^5$ bars) into the interior \cite{Kaspi2018,galanti2019,Galanti2020}, maintaining a cylindrical orientation along the direction of the axis of rotation \cite{Kaspi2023}.}

Deep, convection-driven 3D simulations driven by internal heating
can also reproduce the banded structure of zonal jets through momentum
transport by Reynolds stresses, and laboratory experiments have generally supported these findings \cite{lemasquerier2021}. Studies indicate that the equatorial
jets can form in either direction depending on the dominance of Coriolis
or buoyancy forces (i.e., Rayleigh number and Ekman number, see Materials and Methods) \cite{Aurnou2007,Kaspi2009,gastine2013}.
While this explanation is effective for understanding different stellar
convection model results (the transition between solar and anti-solar
differential rotation states) \cite{gastine2013,featherstone2016,hindman2020,camisassa2022},
it may not apply to gas giants. {The four Jovian planets exhibit similar rotation rates, with the ice giants rotating approximately 1.5 times slower than the gas giants. The radii of the ice giants are about one-third those of the gas giants, and with comparable jet speeds, this results in a planetary Rossby number approximately twice as large for the ice giants. However, the Rossby number remains much less than one for all four planets, indicating the continued dominance of rotation in their dynamical state.} This suggests that the transition proposed by \cite{Aurnou2007} is plausible; however, it likely requires different Rossby numbers, leaving the underlying mechanisms driving such a transition still unclear.

Here, we propose an alternative perspective on the question of equatorial jet direction by utilizing deep convection-driven simulations.
{Our findings show that a subrotation state can also exist under dominant rotational forces, framing this behavior as a bifurcation with two distinct steady solutions. The formation of the equatorial jet is governed by an equivalent mechanism — tilting columnar convection — highlighting notable similarities in the properties of both solutions.} This approach offers a novel
explanation for the distinct equatorial dynamics observed across the
Jovian planets.

\section*{Results}

\subsection*{Theory}

Convection-driven equatorial jet formation has been extensively studied
and modeled in the past [e.g.,~\cite{Busse1976,Busse1986}]. In simple terms, the convection of heat from the planet's interior outward organizes the flow field, correlating between the components of the anomalous flow {(flow velocity that deviates from the averaged value), as long as rotation dominates the momentum balance}. {This organized columnar convection transfers momentum to the mean zonal wind}. 

\begin{figure}[H]
\begin{centering}
\includegraphics[width=0.7\textwidth]{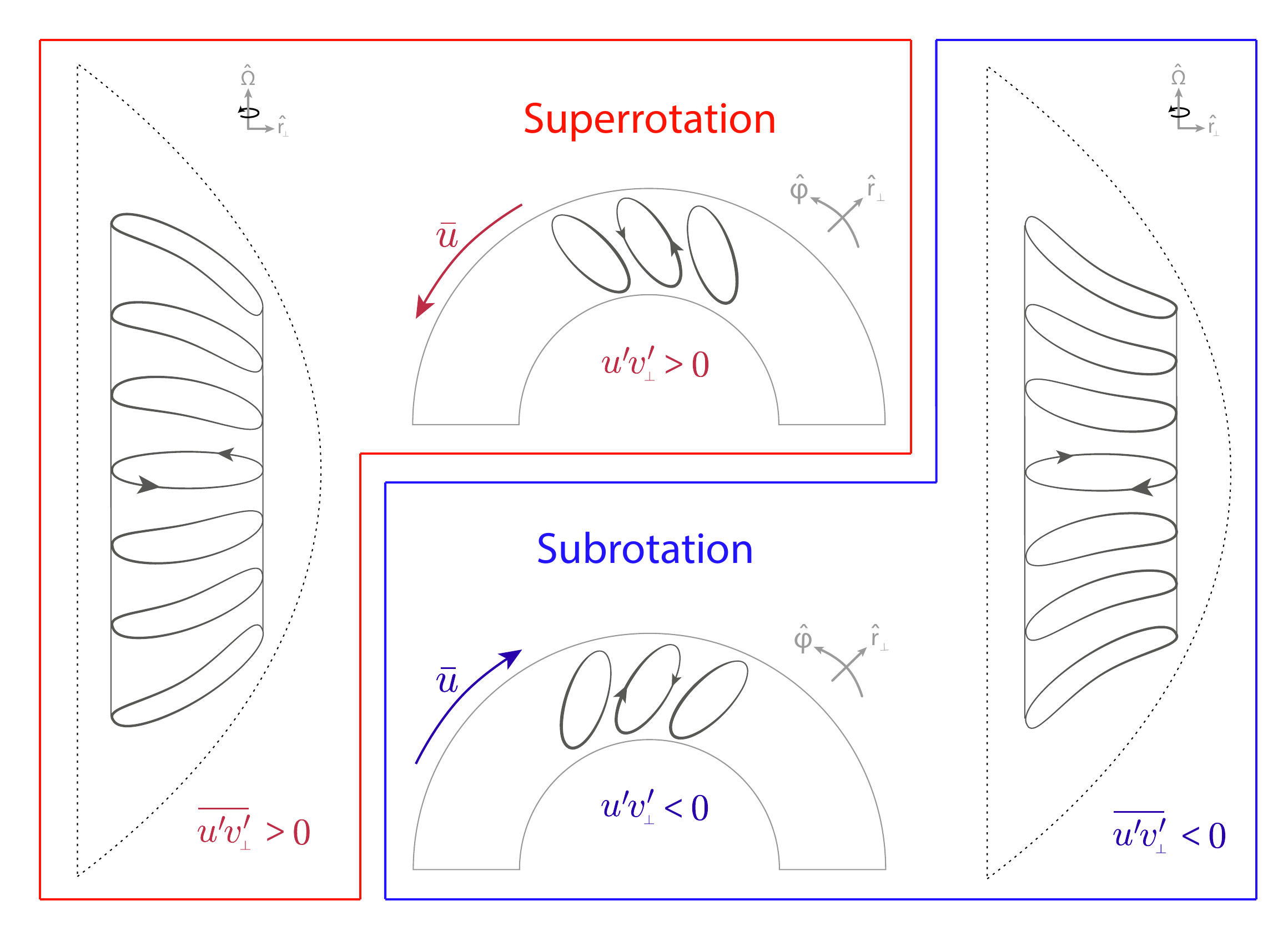}
\par\end{centering}
\caption{\label{fig:Illustration}\textbf{Schematics of tilted convection columns.} The tilt driving
to either outward (superrotation, red) or inward (subrotation, blue) transport
of positive angular momentum with convex boundaries (dashed envelope).
In the subrotation case, the columns have a concave structure while
existing within a convex-boundary system. $\hat{\Omega}$ is the direction of the rotation axis, $\hat{r}_\perp$ is the direction perpendicular to the rotation axis, and $\hat{\varphi}$ is the zonal direction. $\bar{u}$ is the mean flow and $u^\prime v^{\prime}_\perp$ is the eddy momentum flux directed perpendicular to the rotation axis.}
\end{figure}

 {An important aspect of rotating convection is the influence of boundaries that constrain convection cells along the axis of rotation. In the idealized case of a rotating cylindrical annulus, where these boundaries are linear, the circulation within the convective columns (manifested as Reynolds stresses in the momentum balance) facilitates momentum transfer either inward (toward the interior) or outward (toward the outer shell). The direction of this transfer is influenced by the orientation of the columns in the equatorial plane (their tilt), which can develop in either direction (see the equatorial plane cross-sections in Fig.~\ref{fig:Illustration}). This scenario introduces bifurcation, where the tilt can result in two distinct states of mean flow \cite{howard1986}.}
 
  {However, when the boundaries are curved, the tilt of the columns is primarily influenced by the curvature of the sloping boundaries. Intuitively, this can be understood as a divergence in the stretching trend: it vanishes in a linear boundary but reverses in its nature between concave and convex boundaries, dictating a variation during the stretching process. {Note that, in both scenarios, the columns stretch — commonly referred to as the topographic beta effect.}  This variation in curvature results in the formation of eastward or westward shear \cite{Busse1982a,Zhang1992}, leading to a mean zonal flow that corresponds with the tilt (i.e., superrotating or subrotating jets at the equator). Factors such as the density gradient may contribute to similar opposing behaviors \cite{Glatzmaier2009}. The theoretical understanding of this relation between the boundaries and
the direction of the convective driven flow is well established in
previous studies [e.g.,~\cite{Busse1994}]. However, it has not been explored extensively in numerical studies, possibly due to the unconventional nature of the concave columnar structure.}

In spherical
geometries as on planets, only positive momentum injection toward
the outer shell is sustained due to convex-curvature constraints (Fig.~\ref{fig:Illustration},
red). In a hypothetical scenario where a convection column does not
reach the outer boundary, a reverse tilt can develop, manifesting
concave columnar structure, sustaining convection cells and leading
to equatorial subrotation (Fig.~\ref{fig:Illustration}, blue).

\subsection*{Model Formulation}

To simulate the equatorial jet of the Jovian planets, we employ the
open-source, 3D Rayleigh convection model \cite{Featherstone2022}.
The set of equations follow the general formulation used in the hydrodynamics
benchmarks [e.g.,~\cite{Jones2011}]. See Materials and Methods
for more information.

The momentum equation can be written as

\begin{equation}
\frac{\partial\mathbf{u}}{\partial t}+\mathbf{u}\cdot\nabla\mathbf{u}+2\Omega\times\mathbf{u}=\frac{\mathbf{g}S}{C_{p}}-\nabla\left(\frac{p'}{\bar{\rho}}\right)+\frac{1}{\bar{\rho}}\nabla\cdot\mathbf{D},\label{eq:momentum-3}
\end{equation} where $\bar{\rho}\left(r\right)$ is the background-state density,
$\mathbf{u}$ is the velocity vector, $t$ is time, ${\Omega}$ is the
planetary rotation rate, $p'$ is the dynamical pressure, $\mathbf{g}\left(r\right)$
is the gravitational acceleration, $S$ is the entropy and $c_{p}$
is the specific heat capacity at constant pressure. Lastly, $\mathbf{D}$
represents the stress tensor.

The zonally-averaged zonal momentum equation, using Reynolds decomposition
for the velocity vector ($\mathbf{u}=\bar{\mathbf{u}}+\mathbf{u}'$,
where the bar is the zonal average term and prime denotes deviations
from this average, 'eddy') gives:

\begin{equation}
\frac{\partial\bar{u}}{\partial t}+\bar{\rho}\left(\bar{u}\cdot\nabla\mathbf{\bar{u}}+\overline{u^{\prime}\cdot\nabla\mathbf{u}^{\prime}}-2\Omega\sin{\theta}\bar{v}+2\Omega\cos{\theta}\bar{w}\right) ={\rm F},\label{eq:momentum-leading-order}
\end{equation} where the velocity vector $\mathbf{u}$ can be decomposed to $u$
in the zonal direction, $v$ in the meridional direction, and $w$
in the radial direction, $\theta$ is the latitude, and ${\rm F}$ is the viscosity force (see
Materials and Methods for full derivation in spherical coordinates).
The acceleration of the zonal wind is driven by advection of the mean
flow, Reynolds stresses {(hereafter referred to as eddy momentum flux $\overline{u^\prime v^\prime}$ and $\overline{u^\prime w^\prime}$), see Eq.~\ref{eq:momentum-spherical-coord}}, and Coriolis
forces, balanced by viscosity. This equation effectively describes
the fluid motion that generates equatorial-dominant zonal jets, powered
by convection columns oriented {parallel} to the axis of rotation
\cite{Kaspi2009,duer2024}. 

\subsection*{Superrotation and subrotation in equivalent conditions\label{subsec:The-transition-from}}

\begin{figure}[H]
\begin{centering}
\includegraphics[width=1\textwidth]{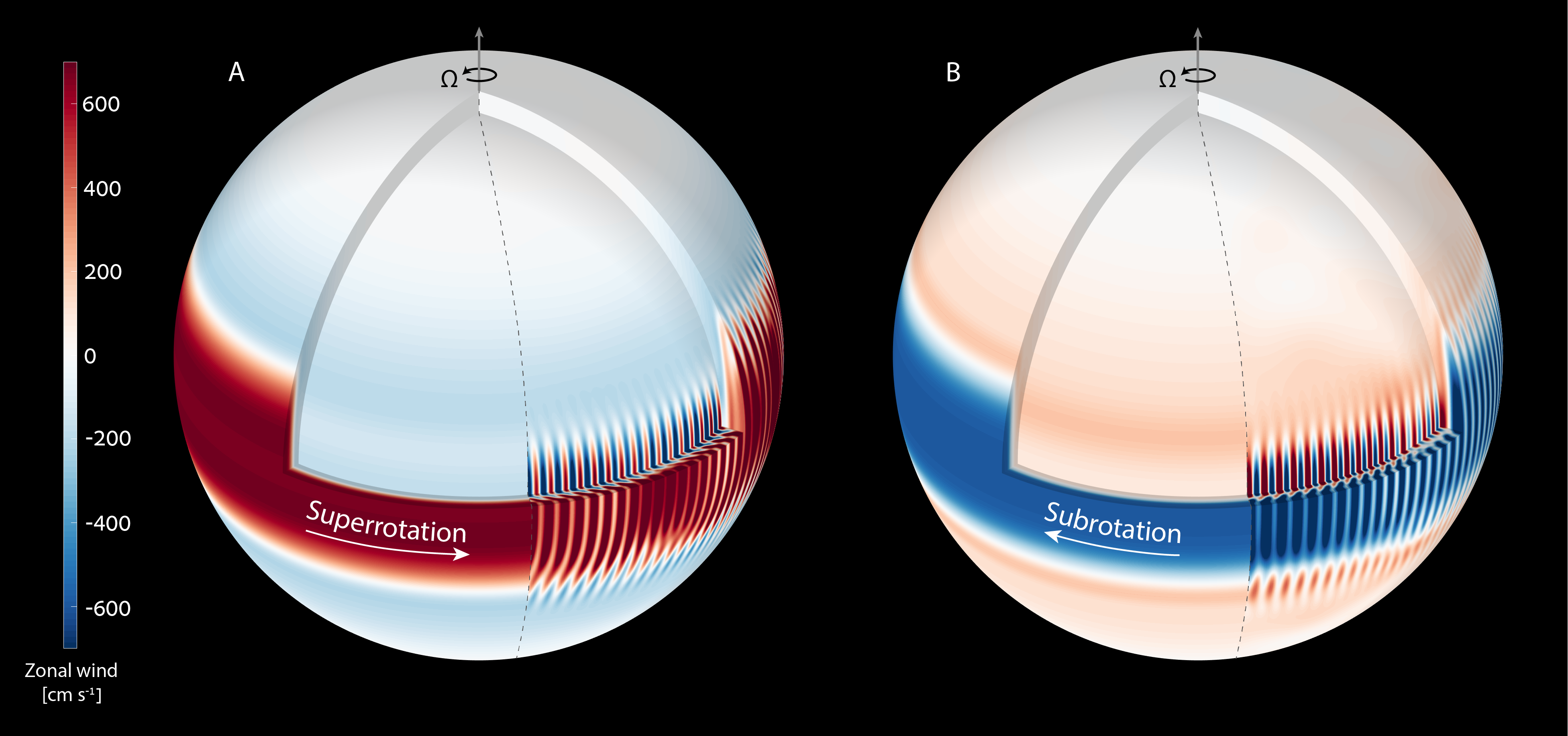}
\par\end{centering}
\caption{\label{fig:3D}\textbf{Results from two simulations with identical physical
parameters, initiated from rest with random entropy noise.} (A) Zonally averaged zonal
wind $\left[{\rm cm\,s^{-1}}\right]$ (left side of the sphere) and
zonal wind (right side of the sphere) of a convection-driven simulation,
showing a superrotating equatorial jet flanked by two retrograde higher-latitude
jets. The right side also displays convection columns tilted in the
prograde direction. Outer shell is shown for $r=r_{o}$ and inside
the slice values are shown for $r=r_{i}$ ($\mu=\frac{r_{i}}{r_{o}}=0.92$).
(B) Similar setup but with reversed patterns: the equatorial jet is
subrotating, the higher-latitude jets are in the prograde direction,
and the convection columns are tilted in the retrograde direction.}
\end{figure}

Keeping the physical parameters constant, {we explore two cases, starting
from rest with random initial entropy noise}. The normalized inner radius ($\mu=\frac{r_{i}}{r_{o}}$, where the subscripts i and o denote the inner and outer boundaries, respectively)
is set to approximately 0.92, comparable to the convective envelopes
of Jupiter, Uranus, and Neptune. {We chose a regime near the onset of rotating convection, namely the weakly non-linear regime, where wave properties and force balances can be reliably estimated and resolved. To achieve this, the non-dimensional control parameters were set to: $Pr=3$, $Ek=7.5\times10^{-4}$, and $Ra^*=0.0132$. It is important to acknowledge that actual planets are expected to exhibit substantially higher levels of turbulence, conditions that are challenging to achieve in numerical simulations \cite{Showman2011}.} All other parameters controlling
the simulations are detailed in the Materials and Methods section.
In Case (a), we observe the well-studied equatorial superrotation driven
by convection (Fig.~\ref{fig:3D}A). The right side of the figure
shows a snapshot of the zonal wind, illustrating the columnar structure
and prograde tilt. The left side presents the zonally-averaged zonal
wind, confirming the superrotation. In Case (b), the scenario is almost
a mirror image of Case (a) (Fig.~\ref{fig:3D}B). The right side reveals
that the zonal wind tilt is oriented in the retrograde direction while
the left side shows the subrotating zonally-averaged zonal wind. 

As the only difference between the two cases shown in Fig.~\ref{fig:3D}
is the random noise from which the simulations were initiated, it
suggests that the problem is essentially a bifurcation problem with
two distinct stable solutions. To validate this, we repeated the experiment
10 times and conducted 20 additional experiments at slightly different
$\mu$ values (approximately 0.91 and 0.93), while keeping other parameters
constant (see Materials and Methods and tabs. S2 and S3). Out of the
30 experiments, 15 resulted in subrotation and 15 in superrotation
(fig.~S4), highlighting the similar likelihood
of achieving either solution. However, this distribution is not strictly half and half, but rather reflects a statistical tendency.

\begin{figure}[H]
\begin{centering}
\includegraphics[width=1\textwidth]{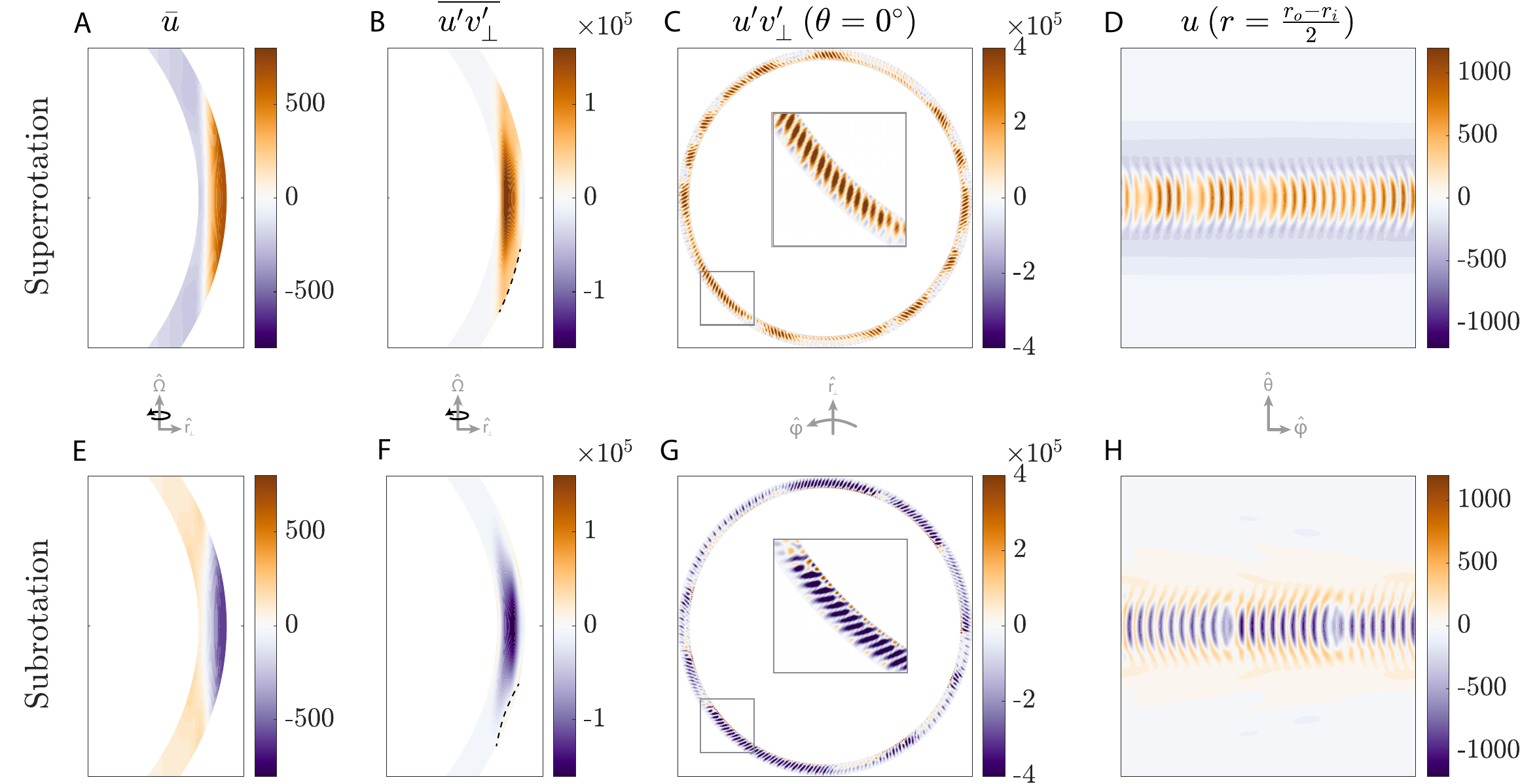}
\par\end{centering}
\caption{\label{fig:Comparision}\textbf{Zonal wind and eddy fluxes in superrotation and subrotation.} Superrotation
in the upper panels and subrotation in the lower panels. (A, E) Zonally-averaged
zonal wind $\left[{\rm cm\,s^{-1}}\right]$ (shown are latitudes -30\textdegree{}
to 30\textdegree ), (B, F) Zonally-averaged eddy fluxes in the direction
perpendicular to the axis of rotation $\left[{\rm cm^{2}\,s^{-2}}\right]$, where the black dashed lines emphasize the curvature of the cylinder at the bottom, (C, G) Snapshot
of eddy fluxes in the equatorial plane $\left[{\rm cm^{2}\,s^{-2}}\right]$,
and (D, H) Snapshot of the zonal wind velocity at mid-shell depth
$\left[{\rm cm\,s^{-1}}\right]$. Both simulations have identical
physical parameters and were initiated from random noise. The geometric orientation of the different panels is shown between the upper and lower rows.}
\end{figure}

Examining the eddy momentum flux terms perpendicular to the axis of rotation
($u^{\prime}v_{\perp}^{\prime},$ which at the equator is equivalent
to $u^{\prime}w^{\prime}$, Eq.~\ref{eq:perp-def}) reveals their columnar structure and orientation.
In the superrotation case, the columns extend to the spherical boundary
of the domain (Fig.~\ref{fig:Comparision}B), and tilt in the prograde
direction ($u^{\prime}v_{\perp}^{\prime}>0$, pumping momentum outwards,
Fig.~\ref{fig:Comparision}C), as predicted by potential vorticity
conservation theory - a conserved quantity that combined the effects of circulation, rotation and stratification \cite{Vallis2006}). In the subrotation case, the columns terminate
before reaching the boundary, exhibiting a concave structure (Fig.~\ref{fig:Comparision}F)
and tilt in the retrograde direction ($u^{\prime}v_{\perp}^{\prime}<0$,
pumping momentum inwards, Fig.~\ref{fig:Comparision}G), a phenomenon
not previously demonstrated within convex boundaries. The stability
of the subrotation and concave columns within the spherical geometry
supports the scenario depicted in Fig.~\ref{fig:Illustration} (blue).

Examining the full momentum budget (Eq.~\ref{eq:momentum-leading-order}
and Eq.~\ref{eq:momentum-spherical-coord}) reveals similarities
between the two cases in terms of the leading order terms (fig.~S1
and fig.~S2). In both cases, the Coriolis force terms and eddy momentum
flux convergence terms are balanced by the viscosity in the model
at steady state. This balance is expected in rotation-dominant convection
models [e.g.,~\cite{Kaspi2009,Jones2009,gastine2013,duer2023,duer2024}].
{In the subrotation case, the columns are truncated before reaching
the domain's geometrical boundary, causing the Taylor--Proudman constraint
to break near the boundaries, exhibiting a sinusoidal pattern in the cylindrical direction, and revealing vertical variations along
the axis of rotation. Theoretical studies often assume that the Taylor--Proudman
constraint holds perfectly, which may explain why this solution has not been
suggested in the past. In practice, though the Taylor–Proudman constraint is not strictly maintained, the length scale is still long enough to almost satisfy the Taylor-Proudman theorem, and the flow remains aligned with the rotation axis, indicating that
rotation still dominates the momentum balance \cite{Kaspi2009}.}

Interestingly, the onset of the two cases appears identical (fig.~S6),
with the solutions beginning to drift towards their final steady states
after approximately 30 rotations. Prior to this point, the outcome remains indeterminate. Analysis of the eddy fluxes at various
depths over time (fig.~S6~A,B,C) highlights
the temporal variability of the columns, revealing a wave-like structural
pattern. At mid-shell depth, a similar analysis shows the temporal
and azimuthal propagation of these waves (fig.~S6~D,E).
In the superrotation case, the waves propagate in the prograde direction,
while in the subrotation case, they propagate in the retrograde direction,
underscoring the antisymmetry between the two scenarios. Another way
to visualize the wave properties is through a frequency-wavenumber
plot (fig.~S7~A,B). {The phase speed of the waves can be directly deduced from the trends in these plots, resulting in absolut values of approximately $1.43\pm 0.07$ ${\rm m\,s^{-1}}$ for the superrotating case
and $1.40\pm 0.05$ ${\rm m\,s^{-1}}$ for the subrotating case (see Materials and Methods and fig.~S7). To verify the nature of these waves, we compare the observed values with those of Thermal Rossby waves, based on the solutions to the linearized equations that describe convection in a rotating annulus \cite{Busse1994,hindman2022}. The theoretical value is given by $c_{{\rm p}}=-\frac{\beta^*}{k^{2}}$ (see Methods), where
$k$ is the typical zonal wavenumber (derived from fig.~S7~C,D for the calculations, specifically $k_{\rm{super}} =2\pi\frac{266}{2\pi R}$ and $k_{\rm{sub}} =2\pi\frac{262}{2\pi R}$), and $\beta^*$ is the topographic beta. At the equator it is $\beta^*(\theta=0^\circ)=\frac{2\Omega}{H}$ (see Materials and Methods). This yields absolute phase speed values of $1.37$ ${\rm m\,s^{-1}}$ and $1.41$ ${\rm m\,s^{-1}}$ for superrotation and subrotation,
respectively. The alignment between the observed and theoretical values not only reinforces the validity of our numerical results, but also emphasizes the origin and antisymmetry of the equatorial waves that dominate the dynamics.}

\subsection*{Bifurcation regime}

\begin{figure}[H]
\begin{centering}
\includegraphics[width=0.8\textwidth]{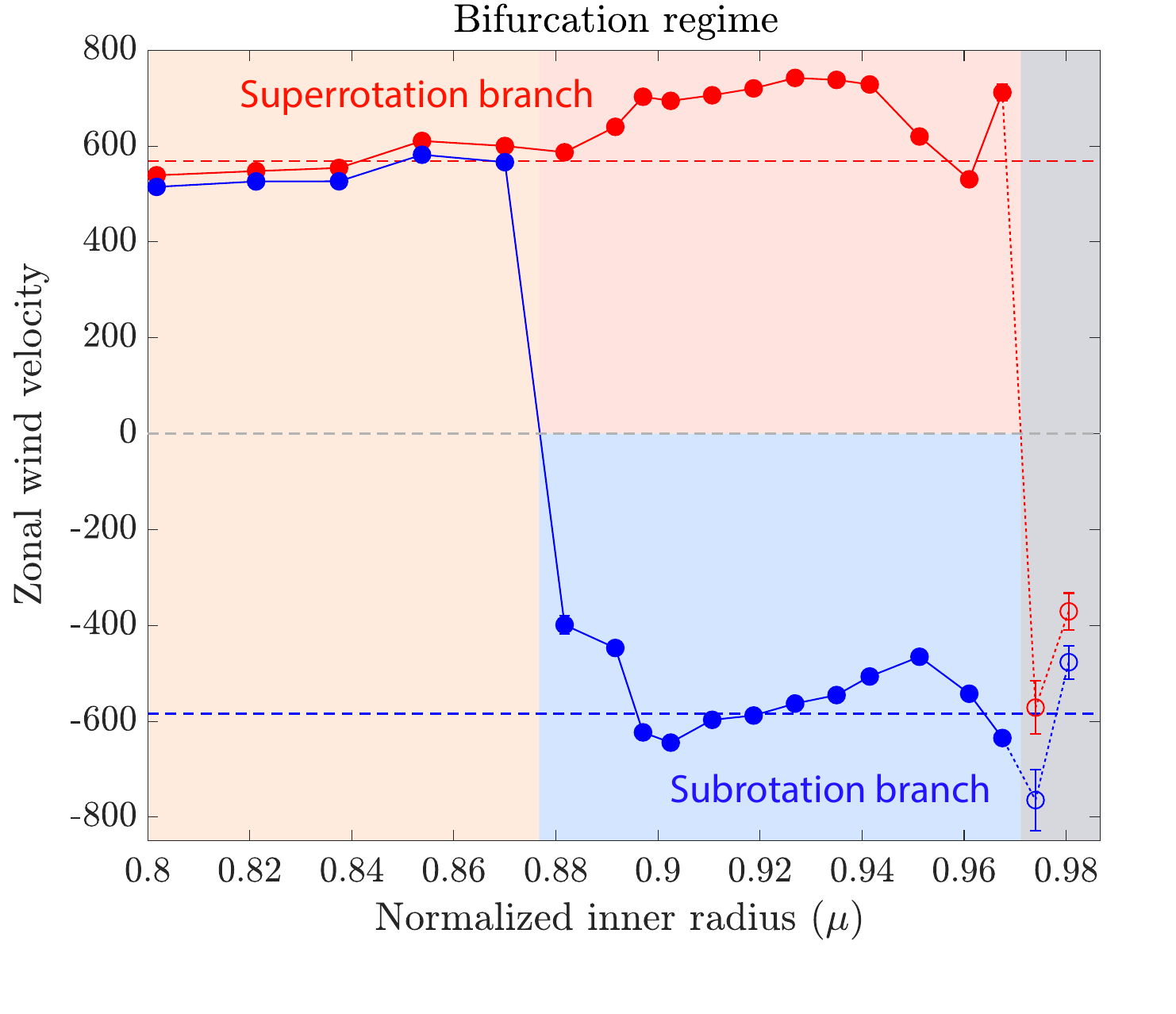}
\par\end{centering}
\caption{\label{fig:bifurcation}\textbf{38 simulations revealing a back-to-back saddle-node
bifurcation.} Two initial conditions are studied: a superrotating case
(red dots, with the red dashed line indicating the initial mean velocity
value) and a subrotating case (blue dots, with the blue dashed line
indicating the initial mean velocity value). For each normalized inner
radius ($\mu=\frac{r_{i}}{r_{o}}$), two simulations were performed
with both sets of initial conditions. All the simulations are in a
statistical steady state, with the mean zonal value and error bars, representing
one SD, are averaged over the last 100 rotations  (mostly notable around the bifurcation nodes, elsewhere the SD is smaller than the circle). The figure reveals four different regions based on $\mu$: A normalized
inner radius smaller than $\mu<0.88$ cannot maintain a concave structure
of columns, resulting in a reversion to the superrotation condition
(orange shade). Depths in the range $\left(0.88<\mu<0.97\right)$
can support both convex (red shade) and concave (blue shade) columnar
structures. {Depths associated with $\mu>0.97$ are presented with open circles and dotted lines as the results are less conclusive (see main text). In this regime (grey shade), only subrotation persists.}}
\end{figure}

Next, we focus on examining the regimes of multiple equilibria. Taking
the two study cases (Fig.~\ref{fig:Comparision}), after reaching
a dynamical steady state, we vary $\mu$ to determine where each solution
holds (tab.~S4). {We adjust other physical parameters while
maintaining the main non-dimensional control parameters and the resolution constant (see Materials and Methods and tab.~S2)}. Starting from the initial conditions shown as dashed horizontal lines in Fig.~\ref{fig:bifurcation} (red for superrotation, blue for subrotation), 38 simulations were performed {(tab.~S4 summarizes the parameters for all the simulations). The resulting steady-state mean zonal wind solutions at the equatorial outer shell are represented by the line and circles.} Each simulation runs
for 2000 rotations (10 viscous diffusion times, see Materials and Methods), and the final equatorial zonal velocity
is presented as the average from the last 100 rotations, along with
one standard deviation (circles, Fig.~\ref{fig:bifurcation}).

The figure illustrates a back-to-back saddle-node bifurcation  \cite{strogatz1994,dijkstra2019}, featuring
two stable branches. The superrotation branch
is the sole stable solution for $\mu<0.88$, indicative of deep domains.
In this regime, the columns are 'aware' of the spherical boundaries,
resulting in a prograde tilt and a superrotating equatorial jet (orange
shade). For mid-values $\left(0.88<\mu<0.97\right)$, both solutions
are possible, suggesting that the final outcome is determined by the
initial conditions. Here, the columns are likely insensitive to the
curvature of the boundaries, allowing for both convex (red shade)
and concave (blue shade) columnar structures to exist, or that a competing
mechanism balances against the convex boundaries, and prevents a preference. In this regime, the relative volume between the cylindrical annulus bounded by the sphere and the curved spherical boundaries is smaller, suggesting a reduced influence of the curved boundaries on the entire column (see
Supplementary Materials). {The subrotation branch continues into the
regime of $0.97<\mu<0.99$, where it becomes the only stable solution
(grey shade). In these relatively shallow domains, the tilt of the columns
is likely influenced by an additional mechanism, such as vigorous
mixing or potential vorticity stretching [e.g.,\cite{Aurnou2007,Kaspi2009}],
inhibiting superrotation. It is also possible that the effective dynamics in this regime exhibit a shallow-layer behavior, resulting in subrotation as commonly observed in models [e.g.,\cite{Chemke2015a,guendelman2025}]. Notably, as the shell thins, the convective structures diminish in size, potentially rendering the resolution too coarse in this thin-shell region. To address this, we employ scaling laws to calculate the minimal resolution required to properly resolve the mode at onset \cite{barik2023}, ensuring adequate simulation resolution for large normalized inner radii. Therefore, for normalized inner radii larger than $\mu=0.95$, we repeat the experiment with double the original resolution (Fig.~S14). The results are generally robust under these changes; however, for very large normalized inner radii ($0.97<\mu<0.99$), achieving sustained simulations necessitates an increase in viscosity, leading to reduced velocity values. Consequently, we cannot unequivocally conclude that the subrotation branch is robust in these very shallow domains ($\mu>0.97$).}

Given the estimated wind penetration depth of the Jovian planets, we can
deduce their characteristics in relation to the bifurcation shown in Fig.~\ref{fig:bifurcation}.
{Our results suggest that planets with mid-range convective envelopes may exhibit either state, while very shallow envelopes show only subrotation. In contrast, planets with a deep dynamical layer are likely to exhibit only superrotation.} Saturn, with its deep convective
envelope \cite{galanti2019}, clearly falls where only superrotation is a stable branch.
In contrast, Jupiter, Uranus, and Neptune lie within regions of two
stable solutions \cite{Kaspi2018,Kaspi2013c}, indicating that their equatorial jets could potentially
orient in either direction. {Since the penetration depth of Uranus
and Neptune is not yet fully established, it's possible that they
could exist in a state where only subrotation is a stable option,
which aligns with the observations.} {It is important to note that the
$\mu$ values in this figure might not correspond exactly to actual
planets, as the results are model-dependent. Moreover, we operate in the weakly non-linear regime of turbulent convection, an idealized scenario for rotating turbulence. Therefore, while the results shown in Fig.\ref{fig:bifurcation} demonstrate the potential for both superrotation and subrotation to occur at large normalized inner radius, further work is needed to directly connect these findings with specific observational constraints (such as the depth of the dynamical region). Lastly, to ensure the bifurcation is not confined to the specific parameter set used, we conducted a series of sensitivity tests. Our results indicate the bifurcation persists as long as convection organizes into columns (i.e., supercriticality remains above critical) within a similar convective regime (see Materials and Methods). When the model becomes excessively forced (Fig.~S10), the subrotation branch is no longer sustained. Nonetheless, the bifurcation remains robust within the Boussinesq framework (Fig.~S13), for slightly less viscous cases (Fig.~S12), and at higher $Pr$ numbers (Fig.~S9). These robustness tests suggest that the bifurcation phenomenon we have identified may have broader implications for understanding the dynamics of rotating fluids, extending beyond the specific context of giant planets in the Solar System.}

\section*{Discussion}

{This study offers new insights into the convective processes driving the emergence of both superrotation and subrotation in fast-rotating spherical shells.} By exploring two cases with identical physical parameters
starting from random initial noise, we find that both states are achieved
once the system reaches a dynamical steady state, suggesting a bifurcation with two stable solutions. This finding is validated by repeated
experiments. Eddy flux analysis reveals temporal variability in a
wave-like structure, with prograde propagation in superrotation and
retrograde propagation in subrotation, highlighting the inherent antisymmetry
in these dynamics.

Examining the eddy flux terms reveals their columnar structure and
orientation. In the superrotation case, the columns extend to the
spherical boundary and tilt in the prograde direction, consistent
with theoretical arguments [e.g.,~\cite{Busse1994}]. In the subrotation
case, the columns terminate before reaching the boundary, exhibiting
a concave structure and retrograde tilt. The full momentum budget
analysis shows that in both cases, the Coriolis force and eddy momentum
flux convergence are balanced by viscosity at steady state. Despite
the Taylor--Proudman constraint breaking near the boundaries in subrotation,
rotation and eddy fluxes remain dominant in the momentum balance,
demonstrating the stability of these dynamics within spherical geometry.
Frequency-wavenumber analysis supports our findings, providing phase
speeds that align closely with theoretical predictions for Thermal
Rossby waves. These insights enhance our understanding of the mechanisms
driving superrotation and subrotation and offer a robust framework
for future studies in geophysical and astrophysical contexts.

{The bifurcating phase space of zonal wind solutions reveals three distinct dynamical regimes. In deep domains, only superrotation is a stable solution, as the spherical boundaries strongly influence the tilt of the convection columns. In the mid-range, both spherical boundaries and mixing could play comparable roles, allowing for either superrotation or subrotation, depending on the initial conditions. In shallow domains, only subrotation is stable, suggesting that the geometrical boundaries lose control over the resulting tilt and zonal wind. In this shallow case, a competing mechanism might play a role, potentially controlling the direction of the zonal wind shear and subsequently influencing the tilt (reverse causality). For instance, it might exhibit two-dimensional behavior, with potential vorticity mixing around the equator dictating a subrotating equatorial jet and consequently an opposite tilt of the convective columns develops. Note also that, the sensitivity of these results to horizontal resolution introduces a degree of uncertainty that warrants careful consideration for very large aspect ratios.}

Our results could bear implications for the different flow behaviors observed in the Jovian planets. Saturn fits within the stable superrotation branch, while Jupiter, Uranus, and Neptune lie within the region of two possible stable solutions, suggesting that their equatorial jet formation mechanisms could be equivalent. Due to to lack of clear knowledge on the wind penetration depth on Uranus and Neptune, they might also fall within the solely stable subrotation branch.

Other mechanisms have been proposed to explain the distinctions between gas and ice giants, including the transition from buoyancy-dominated to rotation-dominated dynamics [e.g., \cite{Aurnou2007, gastine2013, Kaspi2009}], various forcing or dissipation schemes [e.g., \cite{Liu2010, Lian2008, young2019}], notable differences in magnetic fields — particularly the presence of strong nonaxial components in ice giants \cite{soderlund2020}, and compositional variations that notably affect planetary dynamics \cite{Glatzmaier2009, guillot2022}. {While this study introduces a unique unified mechanism to understand the direction of equatorial jets on giant planets, it is important to acknowledge certain limitations. Our simulations focus on a specific, carefully chosen regime near the onset of rotating convection. This regime allows us to isolate and examine the fundamental dynamics responsible for jet formation; however, it does not fully capture the highly turbulent conditions expected in the atmospheres of real planets; such "true" gas-giant-like regimes remain numerically unstable in current  state-of-the-art 3D rotating convection models \cite{Showman2011}. Furthermore, although we have examined the robustness of the bifurcation across various parameter changes (see Figs.~S9-S14), including more turbulent conditions, a comprehensive exploration of the full parameter space---defined by at least 5 key dimensions---remains computationally prohibitive.}

{Despite these constraints, the identification of a bifurcation leading to opposing jet directions, and its potential to explain the diverse zonal wind structures observed in gas and ice giants, has important implications for understanding atmospheric dynamics of giant planets. The emergence of subrotation from tilted columnar convection has not been previously demonstrated in rotating spherical simulations; in this study, we show that it can arise as a stable solution to the governing equations. We further find that this solution is robust across variations in the normalized inner radius. Our findings underscore the possibility of both superrotation and subrotation arising under identical physical conditions, offering a unified framework to interpret the diverse zonal wind patterns observed on the Jovian planets and shedding light on the fundamental mechanisms behind equatorial jet formation.}

\section*{Materials and Methods}

\subsection*{Angular momentum conserving wind}

{The axial component of specific absolute angular momentum (m) can be defined by:}

\begin{equation}
m = R \cos \theta (\Omega R \cos \theta + u),\label{eq:angular_momentum}
\end{equation}
{therefore, the angular momentum-conserving wind of an air parcel at latitude $\theta$ that moves poleward from rest at the equator can be expressed as [e.g., \cite{Vallis2006}]:}

\begin{equation}
U_m = \Omega R \frac{\sin^2 \theta}{\cos \theta}.\label{eq:superrotation}
\end{equation}
{Since this function varies greatly with latitude, it is clear that only low-latitude winds can be superrotating \cite{imamura2020}. In particular, if the wind speed at the equator is greater than zero, this indicates a state of superrotation.}

\subsection*{Planetary Rossby number}

{The planetary Rossby number is a dimensionless parameter that measures the relative importance of inertial forces to Coriolis (rotation) forces. At the equator, it is given by:}

\begin{equation}
Ro = \frac{U}{ \Omega R}.
\end{equation}
{Since ice giants have similar zonal velocities but rotate at roughly two-thirds the rate of gas giants and have radii about third as large, their Rossby numbers are expected to be approximately twice those of the gas giants.}
\begin{equation}
Ro_{\rm{ice}} = \frac{U}{\frac{3}{2} \Omega \times \frac{1}{3} R}=2Ro_{\rm{gas}}.
\end{equation}
{Nonetheless, all four planets have Rossby numbers that are much less than one, indicating that rotation strongly dominates the force balance.}

\subsection*{Formulation}

\subsubsection*{Governing equations}

The basic equations describing the motion of compressible convection
under the anelastic approximation are (following \cite{Jones2011,Dietrich2018})
\begin{equation}
\frac{\partial\mathbf{u}}{\partial t}-\mathbf{u}\times\mathbf{\omega}=-\frac{\nabla p'}{\bar{\rho}}-\nabla\frac{1}{2}\mathbf{u}^{2}-2\overrightarrow{\Omega}\times\mathbf{u}+\mathbf{F}_{\nu}+\frac{\rho'\mathbf{g}}{\bar{\rho}},\label{eq:momentum}
\end{equation}
\begin{equation}
\nabla\cdot\bar{\rho}\mathbf{u}=0,\label{eq:continuity}
\end{equation}
\begin{equation}
\frac{p'}{\bar{p}}=\frac{\rho'}{\bar{\text{\ensuremath{\rho}}}}+\frac{T'}{\bar{T}},\qquad S=\frac{c_{p}}{\gamma}\left(\frac{p'}{\bar{p}}-\frac{\gamma\rho'}{\bar{\rho}}\right),\label{eq:gas}
\end{equation}
\begin{equation}
\bar{\rho}\bar{T}\left(\frac{\partial S}{\partial t}+\mathbf{u\cdot\nabla}S\right)=\nabla\cdot\kappa\bar{\rho}\bar{T}\nabla S+\Pi+Q_{i}.\label{eq:energy}
\end{equation}
The equations above are the momentum equation as shown by \cite{Chandrasekhar1961}
(Eq.~\ref{eq:momentum}), the continuity equation in an anelastic
format (Eq.~\ref{eq:continuity}), a linearized equation of state, a definition for entropy (Eq.~\ref{eq:gas}) and an evolution
equation for entropy (or temperature) (Eq.~\ref{eq:energy}). The
thermodynamic variables ($p$ - pressure, $T$- temperature and $\rho$
- density) are displayed using Reynolds decomposition, denoted with
an overbar for the spatial mean field (reference state) and primes for deviations
from this mean state. Here we take the deviations to be small relative
to the reference state. In the momentum equation $\omega=\nabla\times\mathbf{u}$
is the vorticity and $\mathbf{F_{\nu}}$ being the viscous force with
constant kinematic viscosity $\nu$ assuming zero bulk viscosity {(no internal friction)}\cite{Jones2011}.
The specific entropy $S$ is defined for perfect gas \cite{Landau1959}
and $\gamma=\frac{c_{p}}{c_{v}}$ where $c_{p}$ is the specific heat
capacity at constant pressure and $c_{v}$ at constant volume. Finally,
in the entropy equation $\kappa$ is the turbulent thermal diffusivity,
the energy flux is $-\kappa\bar{\rho}\bar{T}\nabla S$ \cite{Braginsky1995},
$\Pi$ is viscous heating, and $Q_{i}$ is radially dependent internal
heating \cite{Jones2011}. We evolve the equations above in time
to solve for the time-dependent variables.

\subsubsection*{Dimensionless formulation}

{The hydrodynamic set of equations can be written in a dimensionless
form with a set of control parameters.
The main three control parameters of the simulations are the
modified Rayleigh number $Ra^{*}=\frac{g_{o}\Delta S}{c_{p}\Omega^{2}L}$ (where the 'regular' Rayleigh number is $Ra=\frac{g_{o}\Delta S L^3}{\nu\kappa c_p}$),
the Ekman number $Ek=\frac{\nu}{\Omega L^{2}}$, and the Prandtl number
$Pr=\frac{\nu}{\kappa}$ [e.g.,~\cite{Heimpel2016}], where $L$ is the shell depth, $\Delta S$ is the entropy difference across the shell and $g_o$ is the gravity at the top of the domain. The subscripts $i$ and $o$ denote the inner
and outer boundaries, respectively. In addition, we define the dissipation number $Di=\frac{g_{o}L}{c_{p}T_{o}}=\mu\left({\rm e}^{\frac{N_{\rho}}{n}}-1\right)$, where $n$ is the polytropic index, $N_{\rho}=\ln\left(\frac{\rho_{i}}{\rho_{o}}\right)$
is the number of density scale heights across the shell, and $\mu=\frac{r_{i}}{r_{o}}$
is the normalized inner
radius (or aspect ratio) \cite{Jones2011}. 
Then, the set of equations may be written
as \cite{Heimpel2016}:}

\begin{equation}
\frac{\text{\ensuremath{\partial}}\mathbf{u}}{\text{\ensuremath{\partial}}t}+\mathbf{u}\cdot\nabla\mathbf{u}+2{\rm e}_{z}\times\mathbf{u}=Ra^{*}\frac{r_{o}^{2}}{r}S\hat{r}-\nabla\left(\frac{p'}{\bar{\rho}}\right)+\frac{Ek}{\bar{\rho}}\nabla\cdot\mathbf{D},\label{eq:momentum-non-dim}
\end{equation}

\begin{equation}
\bar{\rho}\bar{T}\left(\frac{\partial S}{\partial t}+\mathbf{u}\cdot\nabla S\right)=\frac{Ek}{Pr}\nabla\cdot\bar{\rho}\bar{T}\nabla S+\frac{EkDi}{Ra^{*}}\Pi+Q_{i},\label{eq:energy-non-dim}
\end{equation}

\begin{equation}
\mathbf{D}=2\bar{\rho}\left(e_{ij}-\frac{1}{3}\nabla\cdot\mathbf{u}\right),\quad e_{ij}=\frac{1}{2}\left(\frac{\partial u_{i}}{\partial x_{j}}+\frac{\partial u_{j}}{\partial x_{i}}\right),\label{eq:Viscous-Stress-Tensor-non-dim}
\end{equation}

\begin{equation}
\Pi=2\bar{\rho}\left(e_{ij}e_{ji}-\frac{1}{3}\left(\nabla\cdot\mathbf{u}\right)^{2}\right).\label{eq:Viscous-Heating-non-dim}
\end{equation}

\subsubsection*{Numerical setup}

To analyze equatorial zonal jets in gas giant atmospheres, we utilize
deep convection-driven simulations using the Rayleigh code \cite{Featherstone2022}.
These simulations can qualitatively replicate the characteristics
observed at the outer boundary of the Jovian planets' equatorial regions.
We examine the domain depth ($\mu=\frac{r_{i}}{r_{o}}$ or $L=r_{o}-r_{i}$)
to characterize the bifurcation problem. {All simulations in this study employ free-slip velocity boundary conditions, which are suitable for the outer boundary of a gaseous planet and may also represent the inner boundary of the atmosphere, assuming the presence of a fluid interior below} [e.g.,~\cite{Jones2011}].
{We employ fixed entropy boundary conditions at both boundaries, a choice that is straightforward and frequently used in benchmark models [e.g., \cite{Jones2011}].}

All the simulations use the same control parameters,
with Jupiter's radius and rotation rate. {Since we employ the anelastic dimensional framework, to maintain constant $Ek,$
$Ra^{*}$ and $Pr$ values while varying the inner shell depth, we scale the dimensional quantities (the viscosity, thermal diffusivity, and entropy gradient)
based on $\zeta$, where $\zeta=1-\frac{7\mu-2.45}{4.55}$ (tabs.~S2 and
S3). This scaling originates from an normalized inner radius (aspect ratio) of $\mu=0.35$, where $\zeta=1$. By taking a specific percentage of this shell ($\mu=0.35$), such as half of it, we achieve $\zeta=0.5$.} The simulations run until they reach a
dynamical steady state ($1000-3000$ rotations), {equivalent to 5-15 viscous diffusion time $(t_\nu = \frac{L^2}{\nu})$}, with all results showing
time-averaged values (excluding snapshots).

{The control parameters were chosen within a regime near the onset of rotating convection (i.e., the weakly nonlinear regime \cite{gastine2016}, the supercriticality of the main experiments (Figs.~\ref{fig:3D},\ref{fig:Comparision}) is approximately 5), where the properties of Thermal Rossby waves are distinctly observable, enabling precise quantification and comparison between different solution types. Within this framework, we can properly estimate the fundamental mechanism responsible 
for equatorial jet formation. Although more turbulent models may more accurately resemble the conditions of real planets, they pose considerable numerical challenges and complicate investigations at large aspect ratios \cite{duer2024}.} Encompassing the physical parameters of the planets listed in tab.~S1 within the Rayleigh-Ekman-Prandtl parameter space, we observe substantial variations between these parameters and the numerical values employed in this study and others \cite{Aurnou2007, Christensen2002}, largely due to computational limitations. Nonetheless, research suggests that the fundamental physics demonstrated in numerical models can still be applicable to actual physical environments \cite{Showman2011}.

\subsubsection*{The zonal momentum equation}

Starting from the zonally-averaged zonal momentum equation (Eq.~\ref{eq:momentum-leading-order}),
considering that the mean momentum fluxes were shown to be small in
the ageostrophic order (deviating from the perfect geostrophic balance, the leading order momentum balance as shown in fig.~S3) \cite{Kaspi2009,duer2023,kaspi2008phd}, we can write

\begin{equation}
\frac{\partial\bar{u}}{\partial t}+\bar{\rho}\left(\overline{u^{\prime}\cdot\nabla\mathbf{u}^{\prime}}-2\Omega\sin\theta\overline{v}+2\Omega\cos\theta\overline{w_{{\rm \text{}}}}\right)=\bar{\rho}\nu\nabla^{2}\bar{u}.\label{eq:momentum-no-mean}
\end{equation}
Using the continuity equation we can rearrange to write the zonal
momentum equation in an anelastic form with spherical coordinates:

\begin{equation}
\begin{array}{c}
\frac{\partial\bar{u}}{\partial t}+\frac{1}{r^{2}\cos\theta}\frac{\partial}{\partial\theta}\left(\bar{\rho}\overline{u^{'}v^{'}}\cos^{2}\theta\right)+\frac{1}{r^{2}}\frac{\partial}{\partial r}\left(\bar{\rho}\overline{u^{'}w^{'}}r^{2}\cos\theta\right)-\bar{\rho}2\Omega\sin\theta\overline{v}+\bar{\rho}2\Omega\cos\theta\overline{w_{{\rm \text{}}}}\\
=\frac{\bar{\rho}\nu}{r^{2}\cos\theta}\frac{\partial}{\partial\theta}\left(\cos\theta\frac{\partial\bar{u}}{\partial\theta}\right)+\frac{\bar{\rho}\nu}{r^{2}}\frac{\partial}{\partial r}\left(r^{2}\frac{\partial\bar{u}}{\partial r}\right).
\end{array}\label{eq:momentum-spherical-coord}
\end{equation}
{The eddy momentum flux in the direction perpendicular to the axis of rotation in an anelastic form with spherical coordinates is }

\begin{equation}
\frac{\partial\overline{u^{\prime}v_{\perp}^{\prime}}}{\partial r_{\perp}}=\sin\left(\theta\right)\frac{1}{\bar{\rho}r\cos^{2}\theta}\frac{\partial}{\partial\theta}\left(\bar{\rho}\overline{u^{'}v^{'}}\cos^{2}\theta\right)+\cos\left(\theta\right)\frac{1}{\bar{\rho}r}\frac{\partial}{\partial r}\left(\bar{\rho}\overline{u^{'}w^{'}}r\cos\theta\right).\label{eq:perp-def}
\end{equation}

At steady state, the primary balance in this equation involves the
two Coriolis forces (figs.~S1,~S2).
When these forces are canceled, the remaining Coriolis terms and the
radial eddy momentum flux convergence term are counterbalanced by
the radial numerical viscosity \cite{duer2023,kaspi2008phd}. Away from the equator, the meridional
eddy momentum flux convergence term becomes substantial, contributing
more in the direction perpendicular to the axis of rotation. This
momentum transfer from the eddy fluxes to the zonal velocity sustains
the equatorial zonal wind in both the superrotation and subrotation cases.
This primary balance holds for the entire columnar structure outside
the tangent cylinder, with contributions from the meridional eddy
flux and viscosity terms in the direction perpendicular to the rotation
axis. This balance also holds in the midlatitudes,
where solid or parameterized boundaries can represent
viscosity \cite{duer2021}.

\subsection*{Spectral analysis}

To assess the wave-like structure of the columns (visible in Fig.~4~C,D
for superrotation and Fig.~4~G,H for subrotation), we conducted a
2D spectral analysis in time and in the zonal direction (fig.~S7~A,B),
and also averaged over a large time span for generality (between day
$300$ and day $900$, where data is taken every $1$ day) (fig.~S7~C,D). {The frequency spectrum was calculated by taking the 2-dimentional (zonal and time dimensions) Fourier transform of the kinetic energy at the equator in the middle of the shell using Matlab's fft2.}
The time averaging (fig.~S7~C,D) reveals three dominant wavenumbers at the equator:
1-30, representing the mean flow and large-scale patterns; $\sim260$,
corresponding to the number of columns; and $\sim130$,
representing the half value (fig.~S7~C,D).
This pattern is evident in Fig.~4~C,G (insets), where columns appear
in front-and-back positions, resulting in a secondary peak at the
half value. The difference between the superrotation and subrotation
cases is minor, hence, the overall trend remains consistent. {Examining the dominant wavenumber at onset (fig.~S6~D,E) reveals that these wavenumbers are consistent from onset to dynamical steady state.}
{The phase velocity of the waves apparent in the Hovm\"{o}ller diagrams (fig.~S6~D,E) can be estimated directly
from the frequency - wavenumber spectra by taking the trend of the
brightest features in the figure. We performed linear fit to these features, by filtering the strongest signals (value$>2\times10^6$) (see insets in fig.~S7~A,B). The resulting values appear in the main manuscript along with one standard deviation of the data. The phase speeds derived from our simulations are valid for the parameters used in our study but should not be interpreted as directly applicable to real planetary conditions, as the simulations are not conducted within the true planetary regime (i.e., $Ek$ and $Ra^*$ numbers are much higher than expectation for real planets \cite{Aurnou2007}).}

{To asses the theoretical phase speed for these conditions, we begin with Taylor-expanding the Coriolis parameter around a latitude $\theta_0$ \cite{Vallis2006}. As the effect of varying the height (topographic beta) is equivalent to the effect of varying the rotation parameter used in meteorology, we use the later for simplicity \cite{Busse1994}.

\begin{equation}
f=2\Omega\sin \theta\approx2\Omega\sin\theta_0+2\Omega(\theta-\theta_0)\cos \theta_0= f_0+\beta y,
\end{equation}
{where $f_0=2\Omega\sin\theta_0$, $\beta=\frac{2\Omega\cos\theta_0}{R}$ and $y=R(\theta-\theta_0)$. In the limit of $\theta_0\xrightarrow{}0$, $\beta y \gg f_0$, hence, }

\begin{equation}
f_{\theta_0 \rightarrow{} 0}=\beta y=\frac{2\Omega}{R}R\theta.
\end{equation}
{This is known as the equatorial beta-plane approximation. Starting from the full dispersion relation found under these conditions \cite{Vallis2006,hindman2022}, we can write:}

\begin{equation}
    \omega^2-c^2k^2-\beta\frac{kc^2}{\omega}=(2m+1)\beta c,
\end{equation}
{where $\omega$ is the dispersion relation, $c$ is the propagation speed, $k$ is the zonal wavenumber, and $m=0,1,2...$.} {For low frequency waves, we can neglect terms with $\omega^2$, and the dispersion relation becomes}
\begin{equation}
    -c^2k^2-\beta\frac{kc^2}{\omega}=(2m+1)\beta c,
\end{equation}
{or in a simpler form}
\begin{equation}
    \omega=\frac{-\beta k}{(2m+1)\beta/c+k^2}.
\end{equation}
{In our case, $k^2\gg (2m+1)\beta/c$. We further recall the use of the topographic beta instead of the 'regular' beta, where $\beta^*=2\Omega/H$, and $H$, the scale height, is $H = 0.25R$, corresponding to the height at
the center of the cylinder for $\mu = 0.92$. Then the dispersion relation and phase velocity are}
\begin{equation}
    \omega=-\frac{\beta^* }{k}, \mathrm{\ \ \ \ } c_p=-\frac{\beta^*}{k^2}.
\end{equation}

\subsection*{Variations in the main control parameters}
{Lastly, we investigate the sensitivity of the bifurcation behavior to variations in the main control parameters. It is important to recognize that we are operating outside the realm of true physical regimes relevant to the planets, meaning there are no definitive 'true' values for these parameters to compare against. This disconnect underscores the need to examine how variations in these parameters influence the dynamical behavior of the system.}

{The experiment closely mirrors that described in Fig.~5, where one parameter (e.g., aspect ratio) was varied while the remaining parameters were held constant at their initial values. The calculation runs were conducted for both super- and sub-rotation cases, starting from the steady states shown in Figs. 3 and 4. Specifically, the simulations from Figs. 3 and 4 were restarted for an additional 5 or 10 viscous diffusion times, during which one or more parameters were adjusted according to the specific experiment detailed in the supplementary materials.} We varied four additional control parameters (the Rayleigh number (fig.~S10), the Ekman number (fig.~S12), the Prandtl number (fig.~S9), and the Density scale height (fig.~S13)), testing both half and double their original values (unless stated otherwise, see Supplementary Text), resulting in 20 additional runs. {Furthermore, as higher aspect ratios require high horizontal resolution to properly solve the onset mode \cite{barik2023}, we repeat the experiment in Fig.~\ref{fig:bifurcation} with higher resolution for $\mu>0.95$ for validity. For aspect ratios $\mu>0.97$, the $Ek$ had to be increased as well.}

Generally, operating near the critical value for convection ensures the bifurcation remains stable, provided the parameters do not cross critical thresholds. Exceeding these thresholds leads to the breakdown of organized convection (wave patterns), resulting in vigorous mixing [e.g.,\cite{Aurnou2007}]. This is evident in several validation tests, including those with low $Pr$ values (Fig.S9, bottom row), low $Ra^*$ values (Fig.S10, bottom row), and larger density variations (Fig.S13, upper row). Moving further into supercriticality yields mixed results: increasing $Ra^*$ leads to the disappearance of the subrotation solution (Fig.S10, upper row), yet decreasing viscosity by a similar factor allows the bifurcation to persist (Fig.S12, middle row), suggesting that the bifurcation's stability is sensitive to the overall parameter combination.


\clearpage 

%
\bibliography{kerensbib} 
\bibliographystyle{sciencemag}

%
%
%
%
%
%


\section*{Acknowledgments}
We would like to thank Yamila Miguel for her invaluable feedback and insights in reviewing this work, and Nick Featherstone for his help with implementing the Rayleigh code. We also extend our appreciation to the three dedicated reviewers whose constructive comments greatly improved the clarity and depth of analysis in this manuscript. K.-D.M. thanks the
Institute for Environmental Sustainability (IES) at the Weizmann Institute of Science for the Next-gen Environmental Sustainability Postdoc award (2024-2025) and the Council for Higher Education in Israel for the CHE/PBC Fellowship for Postdoctoral training abroad for women (2024-2026) for providing personal financial support.
\paragraph*{Funding:}
KDM, NG, EG, and YK acknowledge the support
of the Israeli Space Agency, the Israeli Science Foundation (Grant
3731/21), and the Helen Kimmel Center for Planetary Science at the
Weizmann Institute of Science. KDM acknowledges funding from the European Research Council (ERC) under the European Union’s Horizon 2020 research and innovation programme (grant agreement no. 101088557, N-GINE).
\paragraph*{Author contributions:}
        Conceptualization: KDM, EG, YK;  
	Methodology: KDM, NG, YK;  
	Investigation: KDM, NG;  
	Formal analysis: KDM, ET;
        Visualization: KDM, NG;
        Validation: KDM;
        Software: KDM;
	Supervision: YK;  
        Resources: YK;
        Funding acquisition: YK;
        Project administration:YK;
	Writing—original draft: KDM;  
	Writing—review and editing: KDM, YK, EG, NG;  
\paragraph*{Competing interests:}
The authors declare that they have no competing interests.
\paragraph*{Data and materials availability:}
All data needed to evaluate the conclusions in the paper are present in the paper and/or the Supplementary Materials. The numerical results discussed in this study were derived by applying the Rayleigh convection model to solve the hydrodynamic equations. For a detailed overview of the methodology, please refer to \cite{Featherstone2022}. In this context, we utilized a specific set of parameters and a defined reference state, which are elaborated in the main text and further detailed in the Supplementary materials. This approach allowed us to simulate the dynamics of the planetary systems and analyze the resulting behavior under the specified conditions. The choice of parameters and reference state is critical, as they strongly influence the outcomes and interpretations of the simulations. See Materials and Methods section for further details and discussion.
\paragraph*{Additional citations in Supplementary Materials:}\cite{Ingersoll1993,Pearl1990,Hubbard1991,Pearl1991,Guillot1999b,li2010saturn,li2018less,Kaspi2013a,Martins2012,Hanel1981,milcareck2024}.


\section*{Supplementary materials}
Supplementary Text\\
Figs. S1 to S14\\
Tables S1 to S4



\end{document}